\documentclass [11pt]{article}

\usepackage{doublespace}
\usepackage{amssymb}
\usepackage{amsthm}
\usepackage{verbatim}
\usepackage{delarray}

\newcommand{\ket}[1]{{|#1\rangle}}
\newcommand{\bra}[1]{{\langle#1|}}
\newcommand{\braket}[2]{{\langle#1|#2\rangle}}

\newcommand{\phii}{\ket{\phi_i}}
\newcommand{\tphi}{\ket{\tilde{\phi}_i}}
\newcommand{\tph}{\ket{\tilde{\phi}}}
\newcommand{\tphit}{\bra{\tilde{\phi}_i}}

\newcommand{\tr}{\mbox{Tr}}

\newcommand{\deft}{{\ \stackrel{\triangle}{=}\ }}

\newcommand{\tPhi}{\widetilde{\Phi}}
\newcommand{\hx}{\widehat{X}}

\newcommand{\HH}{{\mathcal{H}}}

\newcommand{\R}{{\mathcal{R}}}
\newcommand{\B}{{\mathcal{B}}}

\newcommand{\U}{{\mathcal{U}}}

\newcommand{\G}{{\mathcal{G}}}
\newcommand{\Q}{{\mathcal{Q}}}
\newcommand{\SSS}{{\mathcal{S}}}

\newcommand{\ie}{{\em i.e., }}
\newcommand{\eg}{{\em e.g., }}

\newtheorem{theorem}{Theorem}

\title{\singlespace A Semidefinite Programming Approach to Optimal Unambiguous
Discrimination of Quantum States}
\author{Yonina C. Eldar\footnote{
The author was with the Research Laboratory of Electronics,
Massachusetts Institute of Technology, Cambridge, MA and is now
with the Technion---Israel Institute of Technology, Haifa 32000,
Israel. E-mail: yonina@ee.technion.ac.il.}}

\date{\today}

\begin{document}

\maketitle

\renewcommand{\thefootnote}{}
\footnotetext{
This work is supported in part by BAE Systems Cooperative Agreement
RP6891 under Army Research Laboratory Grant DAAD19-01-2-0008, by the
Army Research Laboratory Collaborative Technology Alliance through BAE
Systems Subcontract RK78554, and by Texas Instruments through the TI
Leadership University Consortium.}
\renewcommand{\thefootnote}{\arabic{footnote}}

\begin{abstract}
\singlespace
In this paper we consider the problem of unambiguous discrimination
between a set of linearly independent pure quantum states.  We show
that the design of the optimal measurement that minimizes the
probability of an inconclusive result can be formulated as a
semidefinite programming problem.  Based on this formulation, we
develop a set of necessary and sufficient conditions for an optimal
quantum measurement.  We show that the optimal measurement can be
computed very efficiently in polynomial time by exploiting the many
well-known algorithms for solving semidefinite programs, which are
guaranteed to converge to the global optimum.

Using the general conditions for optimality, we derive necessary
and sufficient conditions so that the measurement that results in
an equal probability of an inconclusive result for each one of the
quantum states is optimal. We refer to this measurement as the
{\em equal-probability measurement (EPM)}.  We then show that for
any state set, the prior probabilities of the states can be chosen
such that the EPM is optimal.

Finally, we consider state sets with strong symmetry properties and
equal prior probabilities for which the EPM is optimal.  We first
consider geometrically uniform state sets that are defined over a
group of unitary matrices and are generated by a single generating
vector. We then consider compound geometrically uniform state sets
which are generated by a group of unitary matrices using {\em
multiple} generating vectors, where the generating vectors satisfy a
certain (weighted) norm constraint.

\end{abstract}

\begin{singlespace}
\vspace{.5\baselineskip}\noindent  {\em Index Terms\/}---Quantum
detection, unambiguous discrimination, equal-probability
measurement (EPM), semidefinite programming, geometrically uniform
quantum states, compound geometrically uniform quantum states.
\end{singlespace}

\newpage
\section{Introduction}
\label{sec:intro}

In recent years, research into the foundations of quantum physics
has led to the emerging field of quantum information theory
\cite{BS98it}. Quantum information theory refers to the
distinctive information processing properties of quantum systems,
which arise when information is stored in or retrieved from
quantum states. To convey information using quantum states, we may
prepare a quantum system in a pure quantum state, drawn from a
collection of known states $\{\ket{\phi_i},1 \leq i \leq m\}$. To
detect the information, the system is subjected to a quantum
measurement. If the given states $\ket{\phi_i}$ are not
orthogonal, then no measurement can distinguish perfectly between
them \cite{P95}. A fundamental problem therefore is to design
measurements optimized to distinguish between pure nonorthogonal
quantum states.

We may formulate this problem within the framework of quantum
detection, and seek the measurement that minimizes the probability
of a detection error, or more generally, the Bayes cost
\cite{H76,H73,YKL75,EMV02}. More recently, a different approach to
the problem has emerged, which in some cases may be more useful.
This approach, referred to as unambiguous discrimination of
quantum states, combines  error free discrimination with a certain
fraction of inconclusive results. The basic idea, pioneered by
Ivanovic \cite{I87}, is to design a measurement that with a
certain probability returns an inconclusive result, but such that
if the measurement returns an  answer, then the answer is correct
with probability $1$. Given an ensemble consisting of $m$ states,
the measurement therefore consists of $m+1$ measurement operators
corresponding to $m+1$ outcomes, where  $m$ outcomes correspond to
detection of each of the states and the additional outcome
corresponds to an inconclusive result.

Ivanovic \cite{I87} developed a measurement which discriminates
unambiguously between a pair of nonorthogonal pure states.
The  measurement gives
the smallest possible probability of obtaining an inconclusive
result for unambiguous discrimination,
when distinguishing between two linearly independent
nonorthogonal states with
equal prior probabilities.
This
measurement was then further investigated by Dieks \cite{D88}
and Peres \cite{P88}, and was later extended by Jaeger and Shimony
\cite{JS95} to the
case in which the two states have unequal prior
probabilities.

Although the two-state problem is well developed, the problem of
unambiguous discrimination between multiple quantum states has
received considerably less attention. In \cite{PT98} Peres and Terno
consider unambiguous discrimination between $3$ quantum states.
Chefles \cite{C98} showed that a necessary and sufficient condition
for the existence of unambiguous measurements for distinguishing
between $m$ quantum states is that the states are linearly
independent.  He also proposed a
simple suboptimal measurement for
unambiguous discrimination for which the probability of an
inconclusive result is the same regardless of the state of the
system. Equivalently, the measurement yields an equal probability of
correctly detecting each one of the ensemble states. We refer to such
a measurement as an equal-probability measurement (EPM).  Chefles and
Barnett \cite{CB98} developed the optimal measurement for the special
case in which the state vectors form a cyclic set, \ie the vectors are
generated by a cyclic group of unitary matrices using a single
generating vector, and showed that it coincides with the EPM.  In
their paper, they raise the question of whether or not this is the
only case for which the EPM is optimal.

In this paper we develop a general framework for unambiguous state
discrimination which can be applied to any number of states with
arbitrary prior probabilities.  For our measurement we consider
general positive operator-valued measures \cite{H76,P90},
consisting of $m+1$ measurement operators.  We derive a set of
necessary and sufficient conditions for an optimal measurement
that minimizes the probability of an inconclusive result, by
exploiting principles of duality theory in vector space
optimization.  In analogy to the quantum detection problem,
deriving a closed-form analytical expression for the optimal
measurement directly from these conditions is a difficult problem.
However, our formulation has several advantages. First, it readily
lends itself to efficient computational methods.  Specifically, we
show that the optimal measurement can be found by solving a
standard semidefinite program (SDP) \cite{VB96}, which is a convex
optimization problem.  By exploiting the many well-known
algorithms for solving SDPs \cite{NN94,A91t}, the optimal
measurement can be computed very efficiently in polynomial time.
Since an SDP is convex, it does not suffer from local optimums, so
that SDP-based algorithms are guaranteed to converge to the {\em
global} optimum. Second, although the necessary and sufficient
conditions are hard to solve directly, they can be used to verify
a solution. Finally, the necessary and sufficient conditions lead
to further insight into the optimal measurement. In particular,
using these conditions we derive necessary and sufficient
conditions on the state vectors, so that the EPM minimizes the
probability of an inconclusive result. In contrast with the
general optimality conditions, these conditions can be easily
verified given the state ensemble and the prior probabilities.
Using these conditions we show that for {\em any} set of state
vectors the prior probabilities can be chosen such that the EPM is
optimal.

Based on the necessary and sufficient conditions we develop the
optimal measurement for state sets with  broad symmetry
properties. In particular, we consider geometrically uniform (GU)
state sets \cite{F91,EF01,EMV02s} defined over a group of unitary
matrices. For such state sets we show that the optimal measurement
is the EPM, and we obtain a convenient characterization of the EPM
that exploits the state symmetries. We then consider {\em compound
GU (CGU)} state sets \cite{EB01,EMV02s} in which the state vectors
are generated by a group of unitary matrices using {\em multiple}
generating vectors. We obtain a convenient characterization of the
EPM in this case, and show that when the generating vectors
satisfy a certain constraint,  the EPM is optimal.

The paper is organized as follows.  After a statement of the problem
in Section~\ref{sec:ud}, in Section~\ref{sec:sdpf} we derive the
necessary and sufficient conditions for the optimal measurement that
minimizes the probability of an inconclusive result, by formulating
the problem as an SDP. In Section~\ref{sec:epm} we consider the EPM
and derive necessary and sufficient conditions on the state set and
the prior probabilities so
that the EPM is optimal.  Efficient iterative algorithms for
minimizing the probability of an inconclusive result which are
guaranteed to converge to the global optimum are considered in
Section~\ref{sec:comp}.  In Sections~\ref{sec:gu} and \ref{sec:cgu} we
derive the optimal measurement for state sets with certain symmetry
properties, and show that the optimal measurement coincides with the
EPM.

\section{Unambiguous Discrimination of Quantum States}
\label{sec:ud}

Assume that a quantum system is prepared in a pure quantum state
drawn from a collection of given states $\{ \ket{\phi_i},1 \leq i
\leq m \}$ in  an $r$-dimensional complex Hilbert space $\HH$,
with $r \geq m$. The states span a subspace $\U$ of $\HH$. To
detect the state of the system a measurement is constructed
comprising $m+1$ measurement operators $\{\Pi_i,0 \leq i \leq m\}$
that satisfy
\begin{equation}
\label{eq:ident}
\sum_{i=0}^m \Pi_i =I_r.
\end{equation}
The measurement operators are constructed so that either the state is correctly
detected, or the measurement returns an inconclusive result.
Thus, each of the operators $\Pi_i,1 \leq i \leq m$ correspond to
detection of the corresponding states $\ket{\phi_i},1 \leq i \leq m$,
and $\Pi_0$ corresponds to an inconclusive result.

Given that the state of the system is $\ket{\phi_i}$, the probability
of obtaining outcome $k$ is
$\braket{\phi_i}{\Pi_k|\phi_i}$. Therefore, to
ensure that each state is either correctly detected or an inconclusive result is
obtained,  we must have
\begin{equation}
\label{eq:zecond}
\braket{\phi_i}{\Pi_k|\phi_i}=p_i\delta_{ik},\quad 1 \leq i,k \leq m,
\end{equation}
for some $0 \leq p_i \leq 1$. Since from (\ref{eq:ident}),
$\Pi_0=I_r-\sum_{i=1}^m \Pi_i$, (\ref{eq:zecond}) implies that
$\braket{\phi_i}{\Pi_0|\phi_i}=1-p_i$, so that given that the
state of the system is $\ket{\phi_i}$, the state is correctly
detected with probability $p_i$, and an inconclusive result is
returned with probability $1-p_i$.

It was shown in \cite{C98} that  (\ref{eq:zecond}) can be satisfied if
and only if the vectors $\ket{\phi_i}$ are linearly independent,
or equivalently, $\dim \U = m$.
 We
therefore make this assumption  throughout the paper. In this case,
we may choose
\begin{equation}
\label{eq:Pi}
\Pi_i=p_i\tphi \tphit\deft  p_iQ_i ,\quad 1 \leq i \leq m,
\end{equation}
where
\begin{equation}
\label{eq:qi}
Q_i=\tphi \tphit,\quad 1 \leq i \leq m,
\end{equation}
and the vectors $\tphi \in \U$ are the {\em reciprocal states}
 associated with the states
 $\phii$\ \ie there are the unique vectors in $\U$ such that
\begin{equation}
\braket{\tilde{\phi}_i}{\phi_k}=\delta_{ik} ,\quad 1 \leq i,k \leq m.
\end{equation}
With $\Phi$ and $\tPhi$ denoting the matrices of columns $\phii$ and
$\tphi$ respectively,
\begin{equation}
\label{eq:tPhi}
\tPhi=\Phi (\Phi^*\Phi)^{-1}.
\end{equation}
Since the vectors $\phii$ are linearly independent, $\Phi^*\Phi$ is always invertible.
Alternatively,
\begin{equation}
\label{eq:tPhia}
\tPhi=(\Phi\Phi^*)^\dagger \Phi,
\end{equation}
so that
\begin{equation}
\label{eq:tPhiv}
\tphi=(\Phi\Phi^*)^\dagger\ket{\phi_i},
\end{equation}
where
$(\cdot)^\dagger$ denotes the {\em Moore-Penrose pseudo-inverse}
\cite{GV96}; the inverse is taken on the subspace
spanned by the columns of the matrix.

We can immediately verify that the measurement operators given by
(\ref{eq:Pi})  satisfy (\ref{eq:zecond}). If $r=m$ so that the
dimension of $\HH$ is equal to the dimension of the space $\U$ spanned
by the vectors $\phii$, then these operators are the unique operators
satisfying (\ref{eq:zecond}). If on the other hand $r>m$, then the
measurement operators are not strictly unique.
Indeed, any measurement operators of
the form
\begin{equation}
\label{eq:Pi2}
\Pi_i=p_iQ_i+\ket{\mu_i}{\bra{\mu_i}},\quad 1 \leq i \leq m,
\end{equation}
where $\ket{\mu_i} \in \U^\perp$, also satisfy (\ref{eq:zecond}).
Since $\ket{\phi_i} \in \U$, $\braket{\phi_i}{\mu_{k}}=0$ for
every $i,k$ so that the measurement operators given by
(\ref{eq:Pi}) and (\ref{eq:Pi2}) lead to the same detection
probabilities $\braket{\phi_i}{\Pi_k|\phi_i}= p_i\delta_{ik}$. We
may therefore assume without loss of generality that the operators
$\Pi_i$ are restricted to $\U$, so that they have  the form given
by (\ref{eq:Pi}).

If the state $\phii$ is prepared with prior probability $\eta_i$,
then the total probability of correctly detecting the state is
\begin{equation}
\label{eq:Pd}
P_D=\sum_{i=1}^m\eta_i \braket{\phi_i}{\Pi_i|\phi_i}=\sum_{i=1}^m\eta_i p_i.
\end{equation}
Our problem therefore is to choose the measurement operators $\Pi_i=p_iQ_i$,
or equivalently the probabilities $p_i \geq 0$, to maximize $P_D$, subject to
the constraint (\ref{eq:ident}).
We can express this constraint directly in terms of the probabilities $p_i$ as
\begin{equation}
\label{eq:const}
\sum_{i=1}^m\Pi_i=\sum_{i=1}^m p_i Q_i  \leq I_r.
\end{equation}
Note that (\ref{eq:const}) implies
that $p_i \leq 1$.

\section{Semidefinite Programming Formulation}
\label{sec:sdpf}

We now show that our maximization problem (\ref{eq:Pd}) and
(\ref{eq:const}) can be formulated as a
standard semidefinite program (SDP) \cite{VB96,NN94}, which is a convex
optimization problem. There are several advantages to this
formulation. First, the SDP formulation readily lends itself to
efficient computational methods.
Specifically, by exploiting the many well known
algorithms for solving SDPs \cite{VB96}, \eg interior point
methods\footnote{Interior point methods
are iterative algorithms
that terminate once a pre-specified accuracy has been reached.
A worst-case analysis of interior point methods shows that the effort
required to solve an SDP to a given accuracy grows no faster than a
polynomial of the problem size.
In practice, the algorithms behave
much better than predicted by the worst case analysis, and in fact in
many cases the number of iterations is almost constant in the size of
the problem.}  \cite{NN94,A91t}, the optimal measurement can be
computed very efficiently in polynomial time.
 Furthermore,
SDP-based algorithms are guaranteed
to converge to the global optimum.
Second,
by exploiting principles of duality theory in
vector space optimization,
the SDP formulation can be used to derive a set of  necessary and
sufficient conditions for the
probabilities $p_i$ to maximize $P_D$ of (\ref{eq:Pd}) subject to
the constraint (\ref{eq:const}).

We note that recently SDP based methods have been employed in a
variety of different problems in quantum detection and quantum
information \cite{EMV02,JRF02,DPS02,R01,AM01,SZFY02}.

After a description of the general SDP problem in
Section~\ref{sec:sdp}, in Section~\ref{sec:udsdp} we show that our
maximization problem can be formulated as an SDP. Based on this
formulation, we derive a set of necessary and sufficient
conditions on the measurement operators, or equivalently, the
probabilities $p_i$, to minimize the probability of an
inconclusive result. Although in general obtaining a closed form
analytical solution directly from these conditions is a difficult
problem, the conditions can be used to verify whether or not a set
of measurement operators is optimal. Furthermore, these conditions
lead to further insight into the optimal measurement operators. In
particular, in Section~\ref{sec:epm} we use these conditions to
develop necessary and sufficient conditions on the state vectors
and the prior probabilities so that the EPM is optimal.

\subsection{Semidefinite Programming}
\label{sec:sdp}

A standard SDP is the problem of minimizing
\begin{equation}
\label{eq:sdp1}
P(x)=\braket{c}{x}
\end{equation}
subject to
\begin{equation}
\label{eq:sdp2}
F(x) \geq 0,
\end{equation}
where
\begin{equation}
F(x)=F_0+\sum_{i=1}^mx_iF_i.
\end{equation}
Here $\ket{x} \in \R^m$ is the vector to be optimized, $x_i$
denotes the $i$th component of $\ket{x}$, $\ket{c}$ is a given
vector in $\R^m$, and
 $F_i$ are given matrices in the space $\B_n$  of $ n \times n$ Hermitian
matrices\footnote{Although typically in the literature the
matrices $F_i$ are restricted to be real and symmetric, the SDP
formulation can be easily extended to include Hermitian matrices
$F_i$; see \eg \cite{GW01}. In addition, many of the standard
software packages for efficiently solving SDPs, for example the
Self-Dual-Minimization (SeDuMi) package  \cite{S99,PHL02}, allow
for Hermitian matrices.}.

 The
problem of (\ref{eq:sdp1}) and (\ref{eq:sdp2}) is referred to as
the {\em primal problem}. A vector $\ket{x}$ is said to be {\em
primal feasible} if $F(x) \geq 0$, and is {\em strictly  primal
feasible} if $F(x) > 0$. If there exists a strictly feasible
point, then the primal problem is said to be strictly feasible. We
denote the optimal value of $P(x)$ by $\widehat{P}$.

An SDP is a convex optimization problem and can be solved very
efficiently. Furthermore, iterative algorithms for solving SDPs are
guaranteed to converge to the global minimum.
The SDP formulation can also be used to derive necessary and
sufficient conditions for optimality by exploiting principles of
duality theory.
The essential idea is to
formulate a {\em dual problem}
of the form $\max_Z D(Z)$ for some
linear functional $D$
whose maximal value $\widehat{D}$
serves as a
certificate for $\widehat{P}$. That is, for all feasible
values of
$Z \in \B_n$, \ie values of $Z \in \B_n$ that satisfy a certain set of
constraints, and
for all feasible values of $\ket{x}$, $D(Z) \leq P(x)$, so that the
dual problem provides a lower bound on the optimal value of the
original (primal) problem.
If in addition we can establish that $\widehat{P}=\widehat{D}$, then
this equality can be used to develop
conditions of optimality on $\ket{x}$.

The dual
problem associated with the SDP of (\ref{eq:sdp1}) and (\ref{eq:sdp2})
\cite{VB96} is the problem of maximizing
\begin{equation}
\label{eq:dsdp1}
D(Z)=-\tr(F_0Z)
\end{equation}
subject to
\begin{eqnarray}
\label{eq:dsdp2a}
\tr(F_iZ)& = & c_i,\quad 1 \leq i \leq m; \\
\label{eq:dsdp2b}
Z & \geq&  0,
\end{eqnarray}
where $Z \in \B_n$. A matrix $Z \in \B_n$ is said to be {\em dual feasible} if
it satisfies (\ref{eq:dsdp2a}) and (\ref{eq:dsdp2b}) and is
{\em strictly dual
feasible} if  it satisfies (\ref{eq:dsdp2a}) and $Z > 0$.
If there exists a strictly feasible point, then the dual
problem is said to be strictly feasible.

For any feasible $\ket{x}$ and $Z$ we have that
\begin{equation}
P(x)-D(Z)=\braket{c}{x}+\tr(F_0Z)=
\sum_{i=1}^m x_i \tr(F_iZ)+\tr(F_0Z)=
\tr(F(x)Z) \geq 0,
\end{equation}
so that as required, $D(Z) \leq P(x)$. Furthermore, it can be shown \cite{VB96} that if both the primal problem and the dual problem are strictly feasible,
then $\widehat{P}=\widehat{D}$ and $\ket{x}$ is an optimal primal
point if and only if $\ket{x}$ is primal feasible, and there exists a dual feasible $Z \in \B_n$ such that
\begin{equation}
\label{eq:nssdp}
ZF(x)  = 0.
\end{equation}
Equation (\ref{eq:nssdp}) together with (\ref{eq:dsdp2a}), (\ref{eq:dsdp2b}) and (\ref{eq:sdp2})
constitute  a set of necessary and sufficient
conditions for $\ket{x}$ to be an optimal solution to the problem
of (\ref{eq:sdp1}) and (\ref{eq:sdp2}), when both the primal and the dual
are strictly feasible.

If $\widehat{Z}$ maximizes $D(Z)$ so that
$D(\widehat{Z})=\widehat{D}$, then
$\ket{x}$ is optimal if and only if  $F(x) \geq 0$ and
$\widehat{Z}F(x)=0$.

\newpage
\subsection{SDP Formulation of Unambiguous Discrimination}
\label{sec:udsdp}

We now show that the unambiguous discrimination problem of
(\ref{eq:Pd}) and (\ref{eq:const}) can be formulated as an SDP.
Denote by
$\ket{p}$ the vector of components $p_i$ and by $\ket{c}$ the vector of
components $-\eta_i$. Then our problem is to minimize
\begin{equation}
\label{eq:cost} P(p)=\braket{c}{p},
\end{equation}
subject to
\begin{eqnarray}
\label{eq:primalc}
\sum_{i=1}^mp_iQ_i & \leq & I_r; \nonumber \\
p_i & \geq & 0,\quad 1 \leq i \leq m.
\end{eqnarray}
To
formulate this problem as an SDP, let $F_i,0 \leq i \leq m$ be the
block diagonal matrices defined by
\begin{equation}
\label{eq:F}
F_0=\left[
\begin{array}{cccc}
I_r &  & &\\
& 0  & &  \\
 & & \ddots  & \\
 & & & 0
\end{array}
\right],
F_1=\left[
\begin{array}{cccc}
-Q_1  & & & \\
& 1  & &  \\
& &  \ddots  & \\
& & & 0
\end{array}
\right], \ldots,
F_m=\left[
\begin{array}{cccc}
-Q_m & & &  \\
& 0 & &   \\
& & \ddots  & \\
&  & & 1
\end{array}
\right].
\end{equation}
Then
\begin{equation}
\label{eq:Fp} F(p)= F_0+\sum_{i=1}^mp_iF_i= \left[
\begin{array}{cccc}
I_r-\sum_{i=1}^mp_iQ_i & & &  \\
& p_1 & &   \\
& & \ddots  & \\
&  & & p_m
\end{array}
\right],
\end{equation}
so that the constraint $F(p) \geq 0$ is equivalent to
 $\sum_{i=1}^mp_iQ_i \leq I_r$ and $p_i \geq 0,1 \leq i \leq m$.
Thus the problem of
 (\ref{eq:Pd}) and (\ref{eq:const}) reduces to the SDP
\begin{equation}
\label{eq:sdp}
\min_{p \in \R^m} \braket{c}{p} \mbox{ subject to } F(p) \geq 0,
\end{equation}
where $\ket{c}$ is the vector of components $-\eta_i$ with
$\eta_i$ being the prior probability of $\ket{\phi_i}$, and $F(p)$
is given by (\ref{eq:Fp}).

To derive a set of  necessary and sufficient conditions for optimality
 on $\ket{p}$, we
 use the dual problem formulation of a general SDP (\ref{eq:dsdp1})--(\ref{eq:dsdp2b}) to formulate the dual problem associated with
(\ref{eq:sdp}), which reduces to
\begin{equation}
\label{eq:dsdp}
\max_{X \in \B_r} -\tr(X),
\end{equation}
subject to
\begin{eqnarray}
\label{eq:dsdpca}
\tr(Q_i X)-z_i & = & \eta_i,\quad 1 \leq i \leq m;\\
\label{eq:dsdpcb}
X & \geq & 0; \\
\label{eq:dsdpcc}
 z_i & \geq & 0,\quad 1 \leq i \leq m.
\end{eqnarray}

We can immediately verify that both the primal and the dual problem
are strictly feasible. Therefore it follows that $\ket{p}$ is optimal
if and only if the components $p_i$ of $\ket{p}$ satisfy (\ref{eq:primalc}),
there exists a matrix $X$ and scalars $z_i,1 \leq i
\leq m$ that satisfy (\ref{eq:dsdpca})--(\ref{eq:dsdpcc}), and
\begin{eqnarray}
\label{eq:sn}
X(I_r-\sum_{i=1}^m p_iQ_i) & = & 0;\\
\label{eq:sn2}
z_ip_i& = & 0,\quad 1 \leq i \leq m.
\end{eqnarray}
Note that (\ref{eq:sn}) implies that for the optimal choice of $p_i$,
the largest eigenvalue of
$\sum_{i=1}^mp_iQ_i$ must be equal to $1$. This condition has already
been derived in \cite{C98}.

If $\widehat{X}$ and $\hat{z}_i$ maximize (\ref{eq:dsdp})
subject to (\ref{eq:dsdpca})--(\ref{eq:dsdpcc}), then the optimal
values of $p_i$ can be found by solving (\ref{eq:sn}) and
(\ref{eq:sn2}) with $X=\widehat{X}$, $z_i=\hat{z}_i$.

\newpage
We summarize our results in the following theorem:
\begin{theorem}
\label{thm:ns} Let $\{\ket{\phi_i},1 \leq i \leq m\}$ denote a set
of state vectors with prior probabilities $\{\eta_i, 1 \leq i \leq
m\}$ in an $r$-dimensional Hilbert space $\HH$ that span an
$m$-dimensional subspace $\U$ of $\HH$, let $\{\tphi,1 \leq i \leq
m\}$ denote the reciprocal states in $\U$ defined by
$\braket{\tilde{\phi}_i}{\phi_k}=\delta_{ik}$, and let
$Q_i=\tphi\tphit$.  Let $\Lambda$ denote the set of all ordered
sets of constants $\{p_i,1 \leq i \leq m\}$ that satisfy $p_i \geq
0$ and $\sum_{i=1}^m p_iQ_i \leq I_r$, and let $\Gamma$ denote the
set of $r \times r$ Hermitian matrices $X$ satisfying $X \geq 0$
and scalars $z_i \geq 0,1 \leq i \leq m$ such that
$\tr(Q_iX)-z_i=\eta_i$. Consider the problem $\min_{p_i \in
\Lambda}P(p)$ where $P(p)= -\sum_{i=1}^m \eta_i p_i$ and the dual
problem $\max_{X,z_i \in \Gamma} D(X)$ where $D(X)=-\tr(X)$.  Then
\begin{enumerate}
\item For any $p_i \in \Lambda$ and $X,z_i \in \Gamma$, $P(p) \geq
D(X)$;
\item There is an optimal $\ket{p}$, denoted $\ket{\hat{p}}$, such that
$\widehat{P}=P(\hat{p}) \leq P(p)$ for any $\ket{p} \in \Lambda$;
\item There is an optimal $X$ and optimal $z_i$, denoted $\hx$ and
$\hat{z}_i$, such that
$\widehat{D}=D(\hx) \geq D(X)$ for any $X,z_i \in \Gamma$;
\item $\widehat{P}=\widehat{D}$;
\item A set of necessary and sufficient conditions on $\ket{p}$ to minimize
$P(p)$ is that $p_i \in \Lambda$ and there exists $X,z_i \in
\Gamma$ such that
$X(I_r-\sum_{i=1}^m p_iQ_i)=0$ and $z_ip_i=0,1 \leq i \leq m$.
\item Given $\widehat{X}$ and $\hat{z}_i$ a
set of necessary and sufficient conditions on $\ket{p}$ to minimize
$P(p)$ is that $p_i \in \Lambda$,
$\widehat{X}(I_r-\sum_{i=1}^m p_iQ_i)=0$ and $\hat{z}_ip_i=0,1 \leq i \leq m$.
\end{enumerate}
\end{theorem}

As we indicated at the outset, the necessary and sufficient
conditions given by Theorem~\ref{thm:ns} are in general hard to
solve directly, although they can be used to verify a solution. In
addition, these conditions can be used to gain insight into the
optimal measurement operators. In the next section we will use
Theorem~\ref{thm:ns} to develop necessary and sufficient
conditions on a set of state vectors so that the EPM is optimal.
Contrary to the conditions given by Theorem~\ref{thm:ns}, these
conditions can be easily verified.

\section{Equal-Probability Measurement}
\label{sec:epm}

\subsection{Equal-Probability Measurement}
\label{sec:tepm}

A simple measurement that has been employed for unambiguous state
discrimination is the measurement in
which  $p_i=p,1 \leq i \leq m$. This  measurement
results in equal probability of correctly detecting each of the states.
We therefore refer to this measurement as the
Equal-Probability Measurement (EPM).

To determine the value of $p$,
let $\Phi$ have a singular value decomposition (SVD) \cite{GV96,EF01} of
the form
$\Phi=U\Sigma V^*$ where $U$ is an $r \times r$ unitary matrix,
$\Sigma$ is a diagonal $r \times m$ matrix with diagonal elements
$\sigma_i>0$ arranged in descending order so that $\sigma_1 \geq
\sigma_2 \geq \ldots \geq \sigma_m$, and $V$ is an $m \times m$
unitary matrix. Then from (\ref{eq:tPhi}) it follows that
\begin{equation}
\label{eq:tPhi2}
\tPhi=U (\Sigma^\dagger)^* V^*,
\end{equation}
where $\Sigma^\dagger$ is a diagonal $m \times r$ matrix with diagonal
elements $1/\sigma_i$.
Thus,
\begin{equation}
\sum_{i=1}^mQ_i= \sum_{i=1}^m \tphi \tphit=\tPhi\tPhi^*=
U(\Sigma^\dagger)^* \Sigma^\dagger U^*,
\end{equation}
and the largest eigenvalue of $\sum_{i=1}^mQ_i$ is  equal to
$1/\sigma_m^2$. To satisfy the condition (\ref{eq:sn}) the largest
eigenvalue of $p\sum_iQ_i$ must be equal to $1$, so that
\begin{equation}
p=\sigma_m^2.
\end{equation}
Therefore, our problem reduces to finding necessary and sufficient
conditions on the vectors $\phii$ such that $\Pi_i=\sigma_m^2 Q_i$
minimizes the probability of an inconclusive result.

In the next section we develop conditions under which the EPM is
optimal for unambiguous discrimination. In our development, we
consider separately the case in which $\sigma_m$ has multiplicity
$1$ and the case in which $\sigma_m$ has multiplicity greater than
$1$. We derive a set of necessary and sufficient conditions for
optimality of the EPM in the first case, and sufficient conditions
for optimality in the second case. Two broad classes of state sets
that satisfy these conditions are discussed in
Sections~\ref{sec:gu} and \ref{sec:cgu}.

\subsection{Conditions For Optimality}
\label{sec:epmns}

\subsubsection{Necessary and Sufficient Conditions}

Let $s$ denote the multiplicity of $\sigma_m$ so that
 $\sigma_{m}=\sigma_{m-1}=\cdots=\sigma_{m+s-1}$.
We first consider the case in which $s=1$. In this case to satisfy
(\ref{eq:sn}) and (\ref{eq:dsdpcb}) we must have that
\begin{equation}
X=b \ket{u_m}\bra{u_m},
\end{equation}
where $\ket{u_k}$ are the columns of $U$ and $b \geq 0$. In
addition, since $p_i=p>0$, it follows from (\ref{eq:sn2}) that
$z_i=0,1 \leq i \leq m$ so that from (\ref{eq:dsdpca}),
\begin{equation}
\label{eq:tmpc2} \tr(Q_i X)= b|\braket{\tilde{\phi}_i}{u_m}|^2=
\eta_i,\quad 1 \leq i \leq m.
\end{equation}
Now, from (\ref{eq:tPhi2}) we have that
\begin{equation}
\tphi=U (\Sigma^\dagger)^* \ket{v_i},
\end{equation}
where $\ket{v_i}$ denotes the $i$th column of $V^*$. Substituting
into (\ref{eq:tmpc2}),
\begin{equation}
\frac{b}{\sigma_m^2}|v_i(m)|^2= \eta_i,\quad 1 \leq i \leq m,
\end{equation}
where $v_i(k)$ denotes the $k$th component of $\ket{v_i}$. Since
\begin{equation}
\sum_{i=1}^m |v_i(m)|^2=\sum_{i=1}^m\eta_i=1,
\end{equation}
$b$ must be equal to $\sigma_m^2$.

We conclude that when the multiplicity of $\sigma_m$ is equal to
$1$, the EPM is optimal if and only if $|v_i(m)|^2=\eta_i,1 \leq i
\leq m$, \ie if and only if each of the elements in the  last row
of $V^*$ is equal to the prior probability of the corresponding
state.

\subsubsection{Sufficient Conditions}

We now consider the case in which $s>1$. To derive a set of
sufficient conditions for the EPM to be optimal we construct a
matrix $X$ that satisfies the conditions of Theorem~\ref{thm:ns}.

To satisfy (\ref{eq:sn}) and (\ref{eq:dsdpcb}) we let
\begin{equation}
X=\sum_{k=1}^s b_k \ket{u_{m-k+1}}\bra{u_{m-k+1}},
\end{equation}
with $b_k \geq 0$. Since $p_i=p>0$, it follows from (\ref{eq:sn2})
that $z_i=0,1 \leq i \leq m$ so that from (\ref{eq:dsdpca}), $X$
must satisfy
\begin{equation}
\label{eq:tmpc}
\tr(Q_i X)=\sum_{k=1}^s b_k|\braket{\tilde{\phi}_i}{u_{m-k+1}}|^2=
\eta_i,\quad 1 \leq i \leq m.
\end{equation}
Substituting $\tphi=U (\Sigma^\dagger)^* \ket{v_i}$ into
(\ref{eq:tmpc}), we have that the constants $b_k$ must satisfy
\begin{equation}
\frac{1}{\sigma_m^2}\sum_{k=1}^s b_k |v_i(m-k+1)|^2=
\eta_i,\quad 1 \leq i \leq m,
\end{equation}
where $v_i(k)$ denotes the $k$th component of $\ket{v_i}$.

We conclude that the EPM is optimal if there exists constants $b_i
\geq 0,1 \leq i \leq s$ such that
\begin{equation}
\label{eq:epmns}
\left[
\begin{array}{cccc}
|v_1(m)|^2 & |v_1(m-1)|^2  & \cdots & |v_1(m-s+1)|^2 \\
|v_2(m)|^2 & |v_2(m-1)|^2  & \cdots & |v_2(m-s+1)|^2 \\
\vdots &\vdots  & & \vdots  \\
|v_m(m)|^2 & |v_m(m-1)|^2  & \cdots & |v_m(m-s+1)|^2 \\
\end{array}
\right]
\left[
\begin{array}{c}
b_1 \\
\vdots \\
b_s
\end{array}
\right]=
\left[
\begin{array}{c}
\eta_1 \\
\eta_2 \\
\vdots \\
\eta_m
\end{array}
\right].
\end{equation}

The problem of determining whether there exists a $\ket{b}$ with
components $b_i \geq 0$ such that (\ref{eq:epmns}) is satisfied is
equivalent to verifying whether a standard linear program is
feasible. Specifically, in a linear program the objective is to
minimize a linear functional of the vector $\ket{b}$ of the form
$\braket{d}{b}$ for some vector $\ket{d}$, subject to the
constraints $A\ket{b}=\ket{y}$ and\footnote{The inequality is to
be understood as a component-wise inequality.} $\ket{b}\geq 0$ for
some given matrix $A$ and vector $\ket{y}$. A linear program is
feasible if there exists a vector $\ket{b}$ that satisfies the
constraints \cite{BT97}. Thus we can use standard linear
programming techniques to determine whether a $\ket{b}$ exists
that satisfies (\ref{eq:epmns}), or equivalently, whether given a
set of state vectors with given prior probabilities, the EPM is
optimal.

Note, that given a set of state vectors, we can always choose the
prior probabilities $\eta_i$ so that the EPM is optimal. This
follows from the fact that the matrix in (\ref{eq:epmns}) depends
only on the state vectors. Thus, any set of coefficients $b_i \geq
0$ will give a set of $\eta_i \geq 0$ that satisfy
(\ref{eq:epmns}). The coefficients $\eta_i$ will correspond to
probabilities if  $\sum_i \eta_i=1$. Since $\sum_{i=1}^m
|v_i(k)|^2=1$ for all $k$, $\sum_{i=1}^m \eta_i=\sum_{i=1}^s b_i$,
and any set of coefficients $b_i \geq 0$ such that $\sum_ib_i=1$
will result in a set of probabilities $\eta_i$ for which the EPM
is optimal.

In \cite{CB98} the authors raise the question of whether or not
cyclic state sets with equal prior probabilities are the only
state sets for which the EPM is optimal. Here we have shown that
the EPM can be optimal for {\em any} state set, as long as we
choose the prior probabilities correctly. In Sections~\ref{sec:gu}
and \ref{sec:cgu} we consider state sets with equal prior
probabilities for which the EPM is optimal, generalizing the
result in \cite{CB98}.

We summarize our results regarding the EPM in the following
theorem:
\begin{theorem}
\label{thm:epm} Let $\{\ket{\phi_i},1 \leq i \leq m\}$ denote a
set of state vectors with prior probabilities $\{\eta_i, 1 \leq i
\leq m\}$ in a Hilbert space $\HH$ that span an $m$-dimensional
subspace $\U$ of $\HH$, let $\{\tphi,1 \leq i \leq m\}$ denote the
reciprocal vectors in $\U$ defined by
$\braket{\tilde{\phi}_i}{\phi_k}=\delta_{ik}$, and let
$Q_i=\tphi\tphit$. Let $\Phi=U\Sigma V^*$ denote the matrix of
columns $\ket{\phi_i}$, let $\ket{v_i}$ denote the columns of
$V^*$ and $v_i(k)$ the $k$th component of $\ket{v_i}$, let
$\sigma_1 \geq \ldots \geq \sigma_m$ denote the singular values of
$\Phi$, and let $s$ be the multiplicity of $\sigma_m$. Let
$\Pi_i=\sigma_m^2Q_i$ denote the equal-probability measurement
(EPM) operators. Then,
\begin{enumerate}
\item If $s=1$ then the EPM minimizes the probability of an
inconclusive result if and only if
$|v_i(m)|^2=\eta_i$ for $1 \leq i \leq m$;
\item If $s>1$  then the EPM minimizes the probability of an
inconclusive result if there exists constants $b_i \geq 0,1 \leq i
\leq s$ such that (\ref{eq:epmns}) is satisfied;
\item Given a set of state
vectors, we can always choose the prior probabilities $\eta_i$ so that the EPM
is  optimal. Specifically, $\eta_i$ is given by (\ref{eq:epmns}) where
$b_i$ are arbitrary coefficients satisfying $b_i \geq 0$, and
$\sum_{i=1}^m b_i=1$.
\end{enumerate}
\end{theorem}

Theorem~\ref{thm:epm} provides necessary and sufficient conditions
in the case $s=1$ and sufficient conditions in the case $s>1$ for
the EPM to be optimal, which depend on the SVD of $\Phi$ and the
prior probabilities $\eta_i$. It may also be useful to have a
criterion which depends explicitly on the given states
$\ket{\phi_i}$ and the prior probabilities. Theorem~\ref{thm:epms}
below provides a set of sufficient conditions on the states
$\ket{\phi_i}$ and the prior probabilities $\eta_i$ so that the
EPM is optimal. The proof of the Theorem is given in the Appendix.
In Sections~\ref{sec:gu} and \ref{sec:cgu} we discuss some general
classes of state sets that satisfy these conditions.

\newpage
\begin{theorem}
\label{thm:epms}
Let $\{\ket{\phi_i},1 \leq i \leq m\}$ denote a set of state vectors
with prior probabilities $\{\eta_i, 1 \leq i \leq m\}$ in a
Hilbert space $\HH$ that span an $m$-dimensional subspace $\U$
of $\HH$.  Let $\Phi$ denote the matrix of columns $\ket{\phi_i}$, and
let $q$ denote the number of distinct singular values of $\Phi$.
Then the equal-probability measurement
minimizes the probability of an inconclusive result
if $\braket{\phi_i}{(\Phi\Phi^*)^{t/2-1}|\phi_i}=\eta_ia_t$ for $1 \leq i
\leq m$ and $1 \leq t \leq q$, for some constants $a_t$.
\end{theorem}

\section{Computational Aspects}
\label{sec:comp}

In the general case there is no closed-form analytical solution to
the maximization problem (\ref{eq:cost}) subject to
(\ref{eq:primalc}). However, since this problem is a convex
optimization problem, there are very efficient methods for solving
(\ref{eq:cost}). In particular, the optimal vector $\ket{p}$ can
be computed on Matlab using the linear matrix inequality (LMI)
Toolbox. Convenient interfaces for using the LMI toolbox are the
Matlab packages IQC$\beta$ \cite{MKJR00} and
Self-Dual-Minimization (SeDuMi) \cite{S99,PHL02}. These algorithms
are guaranteed to converge to the global optimum in polynomial
time within any desired accuracy.

The number of operations required for each iteration of a general
SDP where $\ket{x} \in \R^m$ and $F_i \in \B_n$ is $O(m^2n^2)$.
However, the computational load can be reduced substantially by
exploiting structure in the matrices $F_i$. In our problem, these
matrices are block diagonal, so that each iteration requires on
the order of $O(m^4)$ operations \cite{VB96}.

To  illustrate the computational steps
involved in
computing the optimal measurement, we now consider a specific example.

Consider the case in which the ensemble consists of
$3$ state vectors with
equal probability $1/3$, where
\begin{equation}
\label{eq:ex}
\ket{\phi_{1}}=
\frac{1}{\sqrt{3}}
\left[
\begin{array}{r}
1  \\
1\\
1
\end{array}
\right],\,\,\,
\ket{\phi_{2}}=\frac{1}{\sqrt{2}}\left[
\begin{array}{r}
1  \\
1 \\
0
\end{array}
\right],\,\,\,
\ket{\phi_{3}}=
\frac{1}{\sqrt{2}}
\left[
\begin{array}{r}
0  \\
1 \\
1
\end{array}
\right].\,\,\,
\end{equation}
To find the optimal measurement operators, we first find the
reciprocal states $\tphi$. With $\Phi$ denoting the matrix of columns
$\ket{\phi_i}$, we have
\begin{equation}
\tPhi=\Phi (\Phi^*\Phi)^{-1}=\left[
\begin{array}{rrr}
1.73 & 0 & -1.41  \\
-1.73 & 1.41 & 1.41 \\
1.73 & -1.41 & 0 \\
\end{array}
\right],
\end{equation}
and the vectors $\tphi$ are the columns of $\tPhi$.
Next, we form the matrices $Q_i=\tphi\bra{\tilde{\phi}_i}$ which
results in
\begin{equation}
Q_1=3\left[
\begin{array}{rrr}
1 & -1 & 1  \\
-1 & 1 & -1 \\
1 & -1 & 1
\end{array}
\right],\,\,\,
Q_2=\left[
\begin{array}{rrr}
0 & 0 &  0 \\
0 & 2 & -2 \\
0 & -2 & 2
\end{array}
\right],\,\,\,
Q_3=\left[
\begin{array}{rrr}
2 & -2 &  0 \\
-2 & 2 & 0 \\
0 & 0 & 0
\end{array}
\right].
\end{equation}
We can now find the optimal vector $\ket{p}$ using the IQC$\beta$
package on Matlab. To this end we first define the matrices $F_i$
according to (\ref{eq:F}), and define
\begin{equation}
\ket{c}=-\frac{1}{3}
\left[
\begin{array}{r}
1  \\
1 \\
1
\end{array}
\right].
\end{equation}
We then generate the following code, assuming that the matrices
$F_i$ and the vector $\ket{c}$ have already been defined in
Matlab.
\begin{singlespace}
\begin{tabular}{ll}
\texttt{>> abst\_init\_lmi} & \% Initializing
the LMI toolbox\\
\texttt{>> p = rectangular(3,1);} & \% Defining a vector $\ket{p}$ of length
$3$ \\
\texttt{>> }$\mathtt{F=F0;}$ & \% Defining the matrix $F(p)$; here
$\mathtt{Fi}=F_i$ \\
\texttt{>> for i=1:3,} & \\
\texttt{>> }$\mathtt{\quad eval(['W = F'\quad num2str(i)]);}$ & \\
\texttt{>> }$\mathtt{\quad F = F + p(i)*W;}$ & \\
\texttt{>> end} & \\
\texttt{>> }$\mathtt{F>0;}$  & \% Imposing the constraint \\
\texttt{>> lmi\_mincx\_tbx(c'*p);} \hspace*{0.4in} & \% Minimizing
$\braket{c}{p}$ subject to the constraint\\
\texttt{>> P=value(p)} & \% Getting the optimal value of $p$
\end{tabular}
\end{singlespace}
\vspace*{0.1in}
\noindent The optimal vector $\ket{p}$ is given by
\begin{equation}
\label{eq:op}
\ket{p}=\left[
\begin{array}{c}
0 \\
0.17 \\
0.17
\end{array}
\right],
\end{equation}
and the optimal measurement operators
$\Pi_i=p_iQ_i$ are
\begin{equation}
\Pi_1=\left[
\begin{array}{rrr}
0 & 0 & 0  \\
0 & 0 & 0 \\
0 & 0 & 0
\end{array}
\right],\,\,\,
\Pi_2=\left[
\begin{array}{rrr}
0 & 0 &  0 \\
0 & 0.34 & -0.34 \\
0 & -0.34 & 0.34
\end{array}
\right],\,\,\,
\Pi_3=\left[
\begin{array}{rrr}
0.34 & -0.34 &  0 \\
-0.34 & 0.34 & 0 \\
0 & 0 & 0
\end{array}
\right].
\end{equation}

We can now use the necessary and sufficient conditions derived in
Section~\ref{sec:udsdp} and summarized in Theorem~\ref{thm:ns} to
verify that $\ket{p}$ given by (\ref{eq:op}) is the optimal
probability vector. To this end we first form the matrix
$T=I_r-\sum_{i=1}^3\Pi_i$. Using the eigendecomposition of $T$ we
conclude that the null space of $T$ has dimension $1$  and is
spanned by the vector
\begin{equation}
\ket{u}=\left[
\begin{array}{r}
-0.81  \\
0.41 \\
0.41
\end{array}
\right].
\end{equation}
Therefore to satisfy (\ref{eq:sn}) and (\ref{eq:dsdpcb}),
$X$ must be equal to $X=a\ket{u}\bra{u}$ for some $a \geq
0$. Since $p_1=0$ and $p_2,p_3>0$,
(\ref{eq:sn2}) and (\ref{eq:dsdpcc}) imply that $z_2=z_3=0$ and
$z_1 \geq 0$. Therefore, from (\ref{eq:dsdpca}) we must have that
\begin{equation}
\label{eq:tmpce}
\tr(Q_2X)=\tr(Q_3X)=\frac{1}{3},
\end{equation}
and
\begin{equation}
\label{eq:tmpce2}
\tr(Q_1X) \geq \frac{1}{3}.
\end{equation}
To satisfy (\ref{eq:tmpce}) we choose
\begin{equation}
a=\frac{1}{3\braket{u}{Q_2|u}}=0.11.
\end{equation}
With this choice of $a$, $\tr(Q_3X)=1/3$ and
$\tr(Q_1X)=0.89>1/3$,
so that the necessary and sufficient conditions are satisfied.

Now, suppose that instead of equal prior probabilities we assume
that the prior probabilities are $\eta_1=0.6$, $\eta_2=0.2$,
$\eta_3=0.2$. These priors where chosen to be equal the elements
of the last row of $V^*$ where $\Phi=U\Sigma V^*$. Since the
smallest square singular value of $\Phi$, $\sigma_3^2=0.07$, has
multiplicity $1$, (\ref{eq:epmns}) is satisfied and the EPM
consisting of the measurement operators $\Pi_i=pQ_i$ with
$p=0.07$,
 minimizes the probability of
an inconclusive result. As before, we can immediately verify that
this is indeed the correct solution using the necessary and
sufficient conditions of Theorem~\ref{thm:ns}. For this choice of
$\Pi_i$, $T=I_r-p\sum_{i=1}^3 Q_i$, and the null space of $T$ is
spanned by the vector
\begin{equation}
\ket{u}=\left[
\begin{array}{r}
0.68  \\
-0.52 \\
-0.52
\end{array}
\right].
\end{equation}
Therefore $X$ must be equal to $X=a\ket{u}\bra{u}$ for some $a \geq
0$. Since $p_i=p>0$ for all $i$, $z_i=0,1 \leq i \leq 3$ so that we
must have
\begin{equation}
\label{eq:tmpc3}
\tr(Q_1X)=0.6,\,\,\,\tr(Q_2X)=0.2,\,\,\,\tr(Q_3X)=0.2.
\end{equation}
If we choose $a=0.6/\braket{u}{Q_1|u}=0.07$, then (\ref{eq:tmpc3})
is satisfied, and the EPM is optimal.

In the remainder of the paper we use the sufficient conditions of
Theorem~\ref{thm:epms}
to derive the optimal unambiguous measurement for state sets with
certain symmetry
properties. The symmetry properties we consider are quite general,
and include many cases of practical interest.
Specifically, in Section~\ref{sec:gu} we consider geometrically
uniform state sets, and in Section~\ref{sec:cgu} we consider compound
geometrically uniform state sets. It is interesting to note that for
these classes of state sets, the optimal measurement that minimizes
the probability of a detection error is also known explicitly
\cite{EF01,EMV02s}.

\section{Geometrically Uniform State Sets}
\label{sec:gu}

In this section we  consider the case in which the
state vectors $\ket{\phi_i}$ are defined over a
group of
unitary matrices and are generated by a single generating vector. Such a
state set is called  {\em geometrically uniform (GU)} \cite{F91}.
We first obtain a convenient characterization of the EPM in this case
and then show that the EPM is optimal.
This result
generalizes a similar result of Chefles and Barnett \cite{CB98}.

\subsection{GU State Sets}

Let $\G$ be a finite  group of $m$ unitary matrices $U_i$ on
$\HH$. That is, $\G$ contains the identity matrix $I_r$; if $\G$
contains $U_i$, then it also contains its inverse $U_i^{-1} =
U_i^*$;  and the product $U_i U_j$ of any two elements of $\G$ is
in $\G$ \cite{A88}.

A state set generated by $\G$ using a single generating vector
$\ket{\phi}$ is a set $\SSS = \{\ket{\phi_i} = U_i\ket{\phi}, U_i
\in \G\}$. The group $\G$ will be called the \emph{generating
group} of $\SSS$. For concreteness we assume that $U_1=I_r$ so
that $\ket{\phi_1}=\ket{\phi}$. Such a state set has strong
symmetry properties and is called GU. For consistency with the
symmetry of $\SSS$, we will assume equiprobable prior
probabilities on $\SSS$.

Alternatively, a state set is GU if given any two states $\ket{\phi_i}$
and $\ket{\phi_j}$
in the set, there is an
isometry (a norm-preserving linear transformation)
that transforms $\ket{\phi_i}$ into $\ket{\phi_j}$ while leaving the set
invariant \cite{F91}.
Intuitively, a state set is GU if it
``looks the same'' geometrically from any of the states in the set.
Some examples of GU state sets are considered in \cite{F91,EF01}.

We note that in \cite{EF01} a GU state set was defined over
an {\em abelian} group of unitary matrices. Here we are not requiring
the group $\G$ to be abelian.

A cyclic state set is a special case of a GU state set in which
the generating group $\G$ has elements $U_i = Z^{i-1}, 1 \le i \le
m$, where $Z$ is a unitary matrix with $Z^m = I_r$.  A cyclic
group generates a cyclic state set $\SSS =
\{\ket{\phi_i}=Z^{i-1}\ket{\phi},1 \leq i \leq m\}$, where
$\ket{\phi}$ is arbitrary.

Any binary state set $\SSS=\{\ket{\phi_1},\ket{\phi_2}\}$ is a GU
cyclic state set, because it can be generated by the binary group
$\G = \{I_r, R\}$, where $R$ is the reflection about the
hyperplane halfway between the two states. Since $R$ represents a
reflection, $R$ is unitary and $R^2=I_r$.

\subsection{The EPM for GU States}

To derive the EPM for a GU state set with generating group $\G$, we
need to determine the reciprocal states $\tphi$. It was shown in
\cite{EB01,EMV02s} that for a GU state set with generating group $\G$,
$\Phi\Phi^*$ commutes
with each of the matrices $U_i \in \G$.  For completeness we repeat
the argument here.
Expressing $\Phi\Phi^*$ as
\begin{equation}
\Phi\Phi^*=\sum_{i=1}^m\ket{\phi_i}\bra{\phi_i}=
\sum_{i=1}^m U_i\ket{\phi}\bra{\phi}U_i^*,
\end{equation}
we have that for all $j$,
\begin{eqnarray}
\Phi\Phi^* U_j & = & \sum_{i=1}^m
U_i\ket{\phi}\bra{\phi} U_i^*U_j \nonumber \\
& = & U_j \sum_{i=1}^m U_j ^*U_i\ket{\phi}\bra{\phi} U_i^*U_j \nonumber \\
& = & U_j \sum_{i=1}^m U_i\ket{\phi}\bra{\phi} U_i \nonumber \\
& = & U_j\Phi\Phi^*,
\end{eqnarray}
since $\{U_j ^*U_i,1 \leq i \leq m\}$ is just a permutation of $\G$.

If $\Phi\Phi^*$ commutes with $U_j$, then
$T=(\Phi\Phi^*)^\dagger$ also commutes
with $U_j$ for all $j$. Thus, from (\ref{eq:tPhiv}) the reciprocal
states are
\begin{equation}
\tphi=T\ket{\phi_i}=T U_i\ket{\phi}=U_iT\ket{\phi}=U_i \tph,
\end{equation}
where
\begin{equation}
\label{eq:tph}
\tph=T \ket{\phi}=(\Phi\Phi^*)^\dagger \ket{\phi}.
\end{equation}
It follows that the reciprocal states are also GU with generating
group $\G$ and
generating vector $\tph$ given by (\ref{eq:tph}).
Therefore, to compute the reciprocal states for a GU state set all we
need is to compute  the generating vector $\tph$.
The remaining vectors are then obtained by applying the
group $\G$ to $\tph$.
The EPM is then given by the measurement operators
\begin{equation}
Q_i=pU_i\tph\bra{\tilde{\phi}}U_i,
\end{equation}
where $p$ is equal to the smallest eigenvalue of $\Phi\Phi^*$.

\subsection{Optimality of the EPM}

We now show that the EPM is optimal for GU state sets with equal
prior probabilities $\eta_i=1/m$. Since $\Phi\Phi^*$ commutes with
$U_j$ for all $j$, $(\Phi\Phi^*)^a$ also commutes with $U_j$ for
any $a$. Therefore for all $t$,
\begin{equation}
\label{eq:ip}
\braket{\phi_i}{(\Phi\Phi^*)^{t/2-1}|\phi_i}=
\braket{\phi}{U_i^*(\Phi\Phi^*)^{t/2-1}U_i|\phi}=
\braket{\phi}{(\Phi\Phi^*)^{t/2-1}U_i^*U_i|\phi}=
\braket{\phi}{(\Phi\Phi^*)^{t/2-1}|\phi}.
\end{equation}
Since $\braket{\phi_i}{(\Phi\Phi^*)^{t/2-1}|\phi_i}$ does not depend
on $i$, from Theorem~\ref{thm:epms} we conclude that the EPM is optimal.

\newpage
We summarize our results regarding GU state sets in the following theorem:
\begin{theorem}[GU state sets]
\label{thm:gu}
Let $\SSS = \{\ket{\phi_i} =
U_i\ket{\phi}, U_i \in \G\}$
be a geometrically uniform (GU) state set generated
by a finite group $\G$ of unitary matrices, where
$\ket{\phi}$ is an arbitrary state, and let $\Phi$ be the matrix of
columns $\ket{\phi_i}$.  Then the
measurement that minimizes the probability of an inconclusive result
is equal to the equal-probability measurement, and
consists of the measurement operators
\[\Pi_i=p\tphi\bra{\tilde{\phi}_i},\]
where $\{\tphi = U_i\tph, U_i \in \G\}$,
\[\tph=(\Phi\Phi^*)^\dagger \ket{\phi},\]
and $p$ is the smallest eigenvalue of
$\Phi\Phi^*$.
\end{theorem}

\subsection{Example Of A GU State Set}
\label{sec:example}
We now consider an example of a GU state set.

Consider the group $\G$ of $m=4$ unitary matrices $U_i$, where
\begin{equation}
U_1=I_4,\,\,\, U_2=\left[
\begin{array}{rrrr}
1 & 0 & 0 & 0\\
0 & -1 & 0 & 0 \\
0 & 0 & 1 & 0 \\
0 & 0 & 0 & -1
\end{array}
\right],\,\,\, U_3=\left[
\begin{array}{rrrr}
1 & 0 & 0 & 0\\
0 & 1 & 0 & 0 \\
0 & 0 & -1 & 0 \\
0 & 0 & 0 & -1
\end{array}
\right],\,\,\, U_4=U_2U_3.
\end{equation}
Let the state set be $\SSS= \{\ket{\phi_i} = U_i\ket{\phi}, 1 \leq
i \leq 4\}$, where
$\ket{\phi}=1/(3\sqrt{2})[2\,\,\,2\,\,\,1\,\,\,3]^*$, so that
\begin{equation}
\Phi=\frac{1}{3\sqrt{2}}\left[
\begin{array}{rrrr}
2 & 2 & 2 & 2\\
2 & -2 & 2 & -2 \\
1 & 1 & -1 & -1 \\
3 & -3 & -3 & 3
\end{array}
\right].
\end{equation}

From Theorem~\ref{thm:gu} the measurement that minimizes the
probability of an inconclusive result is the EPM. Furthermore, the
reciprocal states $\tphi$ are also GU with generating group $\G$
and generator
\begin{equation}
\tph=(\Phi\Phi^*)^\dagger \ket{\phi}=\frac{1}{4\sqrt{2}}\left[
\begin{array}{r}
3 \\
3 \\
6 \\
2
\end{array}
\right],
\end{equation}
so that $\{\tphi=U_i\tph,1 \leq i \leq 4\}$. Since
\begin{equation}
\Phi\Phi^*=\frac{2}{9}\left[
\begin{array}{rrrr}
4 & 0 & 0 & 0\\
0 & 4 & 0 & 0 \\
0 & 0 & 1 & 0 \\
0 & 0 & 0 & 9
\end{array} \right],
\end{equation}
$p=2/9$ and the EPM measurement operators are
$\Pi_i=(2/9)Q_i=(2/9)U_i\tph\bra{\tilde{\phi}}U_i^*$.

We can now use the necessary and sufficient conditions of
Theorem~\ref{thm:ns} to verify that
$\Pi_i=(2/9)\tphi\bra{\tilde{\phi}_i}$ are indeed the optimal
measurement operators. To this end we first form the matrix
$T=I_r-\sum_{i=1}^4\Pi_i$. Using the eigendecomposition of $T$ we
conclude that the null space of $T$ has dimension $1$  and is
spanned by the vector
\begin{equation}
\ket{u}=\left[
\begin{array}{r}
0  \\
0 \\
1 \\
0
\end{array}
\right].
\end{equation}
Therefore to satisfy (\ref{eq:sn}) and (\ref{eq:dsdpcb}), $X$ must
be equal to $X=a\ket{u}\bra{u}$ for some $a \geq 0$. Since
$p_i=2/9>0,1 \leq i \leq 4$, (\ref{eq:sn2}) and (\ref{eq:dsdpcc})
imply that $z_i=0,1 \leq i \leq 4$. Therefore, from
(\ref{eq:dsdpca}) we must have that
\begin{equation}
\label{eq:tmpcet}
\tr(Q_1X)=\tr(Q_2X)=\tr(Q_3X)=\tr(Q_4X)=\frac{1}{4}.
\end{equation}
To satisfy (\ref{eq:tmpcet}) we choose
\begin{equation}
a=\frac{1}{4\braket{u}{Q_1|u}}=\frac{2}{9}.
\end{equation}
With this choice of $a$, $\tr(Q_2X)=\tr(Q_3X)=\tr(Q_4X)=1/4$, so
that as we expect the necessary and sufficient conditions are
satisfied.

\section{Compound Geometrically Uniform State Sets}
\label{sec:cgu}

In Section~\ref{sec:gu} we showed that the optimal measurement for
a GU state set is the EPM associated with this set. We also showed
that the reciprocal states are themselves GU and can therefore be
computed using a single generator.  In this section, we consider
state sets which consist of subsets that are GU, and are therefore
referred to as {\em compound geometrically uniform (CGU)}
\cite{EB01}. As we show, the reciprocal states are also CGU so
that they can be computed using a {\em set} of generators.  Under
a certain condition on the generating vectors, we also show that
the EPM associated with a CGU state set is optimal.

A CGU state set is defined as a set of vectors
$\SSS=\{\ket{\phi_{ik}},1 \leq i \leq l,1 \leq k \leq r\}$ such that
$\ket{\phi_{ik}}=U_i\ket{\phi_k}$, where the matrices $\{U_i,1 \leq i
\leq l\}$ are unitary and form a
 group $\G$, and the vectors $\{\ket{\phi_k},1 \leq k \leq r\}$  are
the generating vectors.
For consistency with the symmetry of
$\SSS$, we will assume equiprobable prior probabilities on $\SSS$.

A CGU state set  is in general not GU. However, for every $k$, the
vectors $\{\ket{\phi_{ik}},1 \leq i \leq l\}$ are a GU state set
with generating group $\G$. Examples of CGU state sets are
considered in \cite{EB01,EMV02s}.

\subsection{The EPM for CGU State Sets}

We now derive the EPM for a CGU state set with equal prior probabilities.
Let $\Phi$ denote the matrix of columns $\ket{\phi_{ik}}$, where the
first $l$ columns correspond to $k=1$, and so forth.
Then
for a CGU state set with generating group $\G$,  it was shown in
\cite{EB01,EMV02s} that $\Phi\Phi^*$ commutes
with each of the matrices $U_i \in \G$.
If $\Phi\Phi^*$ commutes with $U_i$, then
$T=(\Phi\Phi^*)^\dagger$ also commutes
with $U_i$ for all $i$. Thus, the reciprocal states are
\begin{equation}
\ket{\tilde{\phi}_{ik}}=T\ket{\phi_{ik}}=T
U_i\ket{\phi_k}=U_iT\ket{\phi_k}=U_i \ket{\tilde{\phi}_k},
\end{equation}
where
\begin{equation}
\label{eq:muc}
\ket{\tilde{\phi}_k}=T \ket{\phi_k}=(\Phi\Phi^*)^\dagger \ket{\phi_k}.
\end{equation}
Therefore the reciprocal states are also CGU with generating group $\G$ and
generating vectors $\ket{\tilde{\phi}_k}$ given by (\ref{eq:muc}).
To compute these vectors all we
need is to compute  the generating vectors $\ket{\tilde{\phi}_k}$.
The remaining vectors are then obtained by applying the
group $\G$ to each of the generating vectors.

\subsection{CGU State Sets  With GU Generators}

A special class of CGU state sets  is {\em CGU
state sets  with GU generators} \cite{EB01} in which the generating
vectors $\{\ket{\phi_k},1
\leq k \leq r\}$ are
themselves GU. Specifically, $\{\ket{\phi_k}=V_k \ket{\phi}\}$ for
some generator $\ket{\phi}$,
where the matrices $\{V_k,1 \leq k \leq r\}$ are unitary, and form a
group $\Q$.
Examples of CGU state sets with GU generators are considered in \cite{EMV02s}.

Suppose that $U_i$ and $V_k$ commute up to a phase factor for all
$i$ and $k$ so that $U_iV_k=V_kU_ie^{j\theta(i,k)}$ where
$\theta(i,k)$ is an arbitrary phase function that may depend on
the indices $i$ and $k$. In this case we say that $\G$ and $\Q$
commute up to a phase factor (in the  special case in which
$\theta=0$ so that $U_iV_k=V_kU_i$ for all $i,k$, the resulting
state set is GU \cite{EB01}). Then for  all $i,k$, $\Phi\Phi^*$
commutes with $U_iV_k$ \cite{EB01,EMV02s}. The reciprocal states
$\ket{\tilde{\phi}_{ik}}$ of the vectors $\ket{\phi_{ik}}$ are
therefore given by
\begin{equation}
\ket{\tilde{\phi}_{ik}}= T\ket{\phi_{ik}}=T U_iV_k \ket{\phi}=U_iV_kT
\ket{\phi}=U_i V_k\ket{\bar{\phi}},
\end{equation}
where $\ket{\bar{\phi}}=T \ket{\phi}$.
Thus even though the state set is not in
general GU, the reciprocal states can be computed using
a single generating vector.

 Alternatively, we can express $\ket{\tilde{\phi}_{ik}}$ as
$\ket{\tilde{\phi}_{ik}}=U_i\ket{\tilde{\phi}_k}$ where the generators
$\ket{\tilde{\phi}_k}$ are
given by
\begin{equation}
\label{eq:lsmg}
\ket{\tilde{\phi}_k}=V_k\ket{\bar{\phi}}.
\end{equation}
From (\ref{eq:lsmg}) it follows that the generators
$\ket{\tilde{\phi}_k}$ are GU with
generating group $\Q=\{V_k,1 \leq k \leq r\}$ and generator
$\ket{\bar{\phi}}$.

We conclude that for a CGU state set with commuting GU generators and
generating group $\Q$, the reciprocal states
are also CGU  with commuting GU generators and generating
group $\Q$.

\subsection{The Optimal Measurement for CGU State Sets Satisfying a
Weighted Norm
Constraint}

We now show that if the generating vectors $\ket{\phi_k}$ satisfy
\begin{equation}
\label{eq:cgu_cond}
\braket{\phi_k|(\Phi^*\Phi)^{t/2-1}}{\phi_k}=a_t,\quad 1
\leq k \leq r, 1 \leq t \leq q,
\end{equation}
where $q$ is the number of distinct singular values  of $\Phi$,
then the EPM is optimal.

From Theorem~\ref{thm:epms}
it follows that it is sufficient to
show that (\ref{eq:cgu_cond}) implies
\begin{equation}
\label{eq:cgu_cond2}
\braket{\phi_{ik}|(\Phi^*\Phi)^{t/2-1}}{\phi_{ik}}=a_t,\quad 1 \leq i
\leq l, 1 \leq k
\leq r,1 \leq t \leq q.
\end{equation}
Now,
\begin{equation}
(\Phi^*\Phi)^{t/2-1} \ket{\phi_{ik}}=(\Phi^*\Phi)^{t/2-1} U_i
\ket{\phi_k}=
U_i(\Phi^*\Phi)^{t/2-1}\ket{\phi_k},
\end{equation}
so that
\begin{equation}
\braket{\phi_{ik}|(\Phi^*\Phi)^{t/2-1}}{\phi_{ik}}=
\braket{\phi_{k}|U_i^*U_i(\Phi^*\Phi)^{t/2-1}}{\phi_{k}}=
\braket{\phi_{k}|(\Phi^*\Phi)^{t/2-1}}{\phi_{k}}=a_t,
\end{equation}
establishing (\ref{eq:cgu_cond2}).

For CGU state sets  with GU generators $\{\ket{\phi_k}=V_k
\ket{\phi}\}$ where $V_k \in \Q$ and $\G$ and $\Q$ commute up to a
phase factor, the EPM is optimal. This follows from the fact that
in this case (\ref{eq:cgu_cond}) is always satisfied. To see this,
we first note that $V_k$ commutes with $\Phi\Phi^*$ for each $k$
\cite{EB01}. Therefore for all $k$,
\begin{equation}
\braket{\phi_{k}|(\Phi^*\Phi)^{t/2-1}}{\phi_{k}}=
\braket{\phi|V_k^*(\Phi^*\Phi)^{t/2-1}V_k}{\phi}=
\braket{\phi|V_k^*V_k(\Phi^*\Phi)^{t/2-1}}{\phi}=
\braket{\phi|(\Phi^*\Phi)^{t/2-1}}{\phi}.
\end{equation}

\newpage
We summarize our results regarding CGU state sets in the following theorem:
\begin{theorem}[CGU state sets]
\label{thm:cgu}
Let $\SSS = \{\ket{\phi_{ik}} =
U_i\ket{\phi_k}, 1 \leq i \leq l,1 \leq k \leq r\}$,
be a compound geometrically uniform (CGU) state set generated
by a finite group $\G=\{U_i,1 \leq i \leq l\}$ of unitary matrices and
generating vectors $\{\ket{\phi_k},1 \leq k \leq r\}$,
and let $\Phi$ be the matrix of
columns $\ket{\phi_{ik}}$.
Then the equal-probability measurement (EPM)
consists of the measurement operators
\[\Pi_i=p\ket{\tilde{\phi}_{ik}}\bra{\tilde{\phi}_{ik}},\]
where $\{\ket{\tilde{\phi}_{ik}} = U_i\ket{\tilde{\phi}_k}, 1 \leq i \leq l,1
\leq k \leq r\}$,
\[\ket{\tilde{\phi}_k}=(\Phi\Phi^*)^\dagger \ket{\phi_k},\]
and $p$ is equal to the smallest eigenvalue of $\Phi\Phi^*$.\\
The EPM has the following properties:
\begin{enumerate}
\item If
$\braket{\phi_k|(\Phi\Phi^*)^{t/2-1}}{\phi_k}=a_t$ for $1 \leq k
\leq r, 1 \leq t \leq q$ where $q$ is the number of distinct
eigenvalues of $\Phi\Phi^*$, then the EPM minimizes the
probability of an inconclusive result.
\item If  the generating vectors
$\{\ket{\phi_k}=V_k\ket{\phi},1 \leq k
\leq r\}$ are geometrically uniform
with  $U_iV_k=V_kU_ie^{j\theta(i,k)}$
for all
$i,k$, then
\begin{enumerate}
\item $\ket{\tilde{\phi}_{ik}}=U_iV_k\ket{\bar{\phi}}$ where
$\ket{\bar{\phi}}=(\Phi\Phi^*)^\dagger\ket{\phi}$ so that the
reciprocal states are  CGU with geometrically uniform generators;
\item The EPM is optimal;
\item If in addition $\theta(i,k)=0$ for all $i,k$, then the vectors
$\{\phi_{ik}, 1 \leq i \leq l,1 \leq k \leq r\}$ form a
geometrically uniform state set.
\end{enumerate}
\end{enumerate}
\end{theorem}

\section*{Acknowledgments}

The author wishes to thank Prof. A.\ Megretski and Prof.\ G.\ C.\
Verghese for many helpful discussions on semidefinite programming.

\section*{\centering Appendix}

\section*{Proof of Theorem~\ref{thm:epms}}

In this appendix we prove Theorem~\ref{thm:epms}.

Let $\lambda_i,1 \leq i \leq q$ denote the singular values of $\Phi$ without
multiplicity so that $\lambda_1=\sigma_1$ and $\lambda_q=\sigma_m$, and
let $s_i$ denote the multiplicity of $\lambda_i$.
Define
\begin{equation}
\label{eq:A}
A=\left[
\begin{array}{cccc}
\lambda_1 & \lambda_2  & \cdots & \lambda_q \\
\lambda_1^2 & \lambda_2^2  & \cdots & \lambda_q^2 \\
\vdots &\vdots  & & \vdots  \\
\lambda_1^q & \lambda_2^q  & \cdots & \lambda_q^q \\
\end{array}
\right],
\end{equation}
and
\begin{equation}
\label{eq:X}
H=\left[
\begin{array}{cccc}
\sum_{i=1}^{s_1}|v_1(i)|^2 & \sum_{i=1}^{s_1}|v_2(i)|^2  & \cdots &
\sum_{i=1}^{s_1}|v_2(i)|^2  \\
\sum_{i=1}^{s_2}|v_1(s_1+i)|^2 & \sum_{i=1}^{s_2}|v_2(s_1+i)|^2  & \cdots &
\sum_{i=1}^{s_2}|v_2(s_1+i)|^2  \\
\vdots &\vdots  & & \vdots  \\
\sum_{i=1}^{s_q}\beta_i |v_1(m-s_q+i)|^2 &
\sum_{i=1}^{s_q}\beta_i |v_2(m-s_q+i)|^2  & \cdots &
\sum_{i=1}^{s_q}\beta_i |v_2(m-s_q+i)|^2  \\
\end{array}
\right],
\end{equation}
for some $\beta_i \geq 0$.
Finally, let $N$ be the matrix with $i$th column equal to $\eta_i
\ket{a}$ where $\ket{a}$ is an arbitrary vector.

Now, suppose that $AH=N$. Then $A \ket{h_i}=\eta_i\ket{a}$ where
$\ket{h_i}$ denotes the $i$th column of $H$. Since $A$ is invertible,
this implies that
\begin{equation}
\label{eq:xcond}
\frac{1}{\eta_i}h_i(k)=\frac{1}{\eta_j}h_j(k)\quad 1 \leq i,j \leq m,1
\leq k \leq q.
\end{equation}
For $k=q$, (\ref{eq:xcond}) reduces to (\ref{eq:epmns}). We
therefore conclude that a sufficient condition for the EPM to be
optimal is that $AH=N$ for some $\beta_i \geq 0$. Taking
$\beta_i=1$ for each $i$, we can express $AH$ as
\begin{equation}
\label{eq:H}
AH=\left[
\begin{array}{cccc}
\sigma_1 & \sigma_2  & \cdots & \sigma_m \\
\sigma_1^2 & \sigma_2^2  & \cdots & \sigma_m^2 \\
\vdots &\vdots  & & \vdots  \\
\sigma_1^q & \sigma_2^q  & \cdots & \sigma_m^q \\
\end{array}
\right]
\left[
\begin{array}{cccc}
|v_1(1)|^2 & |v_2(1)|^2  & \cdots & |v_2(m)|^2  \\
|v_1(2)|^2 & |v_2(2)|^2  & \cdots & |v_2(2)|^2  \\
\vdots &\vdots  & & \vdots  \\
|v_1(m)|^2 & |v_2(m)|^2  & \cdots & |v_2(m)|^2  \\
\end{array}
\right] \deft Y.
\end{equation}
Then we have that
\begin{equation}
Y_{tl}=\sum_{i=1}^m\sigma_i^t|v_l(i)|^2=\braket{\phi_l}{(\Phi\Phi^*)^{t/2-1}|\phi_l}.
\end{equation}
Therefore $AH=N$  reduces to the condition that
\begin{equation}
\label{eq:epms}
\braket{\phi_l}{(\Phi\Phi^*)^{t/2-1}|\phi_l}=\eta_la_t,\quad 1 \leq l
\leq m,1 \leq t \leq q,
\end{equation}
for some constants $a_t$.

\newpage
\begin{singlespace}

\end{singlespace}


\end{document}